\documentstyle[12pt]{article}
\begin{document}
\newpage
\pagestyle{empty}
\setcounter{page}{0}
%
\def\CPbar{\hbox{{\rm CP}\hskip-1.80em{/}}}
\def\hfl#1#2{\smash{\mathop{\hbox to 12mm{\rightarrowfill}}
\limits^{\scriptstyle#1}_{\scriptstyle#2}}}
\def\vfl#1#2{\llap{$\scriptstyle #1$}\left\downarrow
\vbox to 6mm{}\right.\rlap{$\scriptstyle#2$}}
\def\ihfl#1#2{\smash{\mathop{\hbox to 12mm{\nearrowfill}}
\limits^{\scriptstyle#1}_{\scriptstyle#2}}}
\def\ibfl#1#2{\smash{\mathop{\hbox to 12mm{\searrowfill}}
\limits^{\scriptstyle#1}_{\scriptstyle#2}}}
\def\diagram#1{\def\normalbaselines{\baselineskip=0pt
\lineskip=10pt\lineskiplimit=1pt}  \matrix{#1}}
\def\adots{\mathinner{\mkern2mu\raise1pt\hbox{.}
\mkern3mu\raise4pt\hbox{.}\mkern1mu\raise7pt\hbox{.}}}

\newfont{\twelvemsb}{msbm10 scaled\magstep1}
\newfont{\eightmsb}{msbm8}
\newfont{\sixmsb}{msbm6}
\newfam\msbfam
\textfont\msbfam=\twelvemsb
\scriptfont\msbfam=\eightmsb
\scriptscriptfont\msbfam=\sixmsb
\catcode`\@=11
\def\Bbb{\ifmmode\let\next\Bbb@\else
  \def\next{\errmessage{Use \string\Bbb\space only in math mode}}\fi\next}
\def\Bbb@#1{{\Bbb@@{#1}}}
\def\Bbb@@#1{\fam\msbfam#1}
\newfont{\twelvegoth}{eufm10 scaled\magstep1}
\newfont{\tengoth}{eufm10}
\newfont{\eightgoth}{eufm8}
\newfont{\sixgoth}{eufm6}
\newfam\gothfam
\textfont\gothfam=\twelvegoth
\scriptfont\gothfam=\eightgoth
\scriptscriptfont\gothfam=\sixgoth
\def\frak{\frak@}
\def\frak@#1{{\fam\gothfam{{#1}}}}
\def\frak@@#1{\fam\gothfam#1}
\catcode`@=12

%
%
%
\def\CC{{\Bbb C}}
\def\NN{{\Bbb N}}
\def\QQ{{\Bbb Q}}
\def\RR{{\Bbb R}}
\def\ZZ{{\Bbb Z}}
\def\cA{{\cal A}}          \def\cB{{\cal B}}          \def\cC{{\cal C}}
\def\cD{{\cal D}}          \def\cE{{\cal E}}          \def\cF{{\cal F}}
\def\cG{{\cal G}}          \def\cH{{\cal H}}          \def\cI{{\cal I}}
\def\cJ{{\cal J}}          \def\cK{{\cal K}}          \def\cL{{\cal L}} 
\def\cM{{\cal M}}          \def\cN{{\cal N}}          \def\cO{{\cal O}}
\def\cP{{\cal P}}          \def\cQ{{\cal Q}}          \def\cR{{\cal R}} 
\def\cS{{\cal S}}          \def\cT{{\cal T}}          \def\cU{{\cal U}}
\def\cV{{\cal V}}          \def\cW{{\cal W}}          \def\cX{{\cal X}}
\def\cY{{\cal Y}}          \def\cZ{{\cal Z}}
\def\qed{\hfill \rule{5pt}{5pt}}
\newtheorem{lemma}{Lemma}
\newtheorem{prop}{Proposition}
\newtheorem{theo}{Theorem}
\newenvironment{result}{\vspace{.2cm} \em}{\vspace{.2cm}}

\rightline{q-alg/9701003} 
\rightline{January 97} 

\vfill

\begin{center}

{\LARGE {
\bf {\sf 
On Generalized Quantum Deformations and Symmetries in Quantum Mechanics
}}}\\[2cm]

\smallskip 

{\large J. Beckers\footnote{E-mail: Jules.Beckers@ulg.ac.be. }}

\smallskip 

\smallskip 

\smallskip

{\em Theoretical and Mathematical Physics,\\ Institute of Physics (B.5),\\
University of Li\`ege,\\
 B-4000 LIEGE 1 (Belgium)}

\end{center}

\smallskip 

\begin{abstract}
Supersymmetric and parasupersymmetric quantum mechanics are now recognized 
as two further parts of quantum mechanics containing a lot of new informations
 enlightening (solvable) physical applications. Both contents are here
 analysed in connection with generalized quantum deformations. In fact, the
 parasupersymmetric context is visited when the order of paraquantization
 $p$ is limited to the first nontrivial value $p = 2$.
\end{abstract}

\vfill 
\vfill

\newpage

\pagestyle{plain}

\section{Introduction}
Recent developments in quantum mechanics are due to simple ideas
 related to the superposition of different types of particles described by
 specific statistics. The simplest context surely is that of supersymmetric
 quantum mechanics (SSQM) subtending the idea of superposition of usual
 (integer spin) $\underline{\hbox{bosons}}$ satisfying the very well known 
Bose-Einstein
 statistics with usual (half integer spin) $\underline{\hbox{fermions}}$ satisfying 
the very well known Fermi-Dirac statistics. Such an idea, first developed by
 Witten \cite{1}, is evidently issued from the rich concept of supersymmetry
 initially exploited in particle physics in the seventies \cite{2}. Its 
recent developments in SSQM are quoted in two review papers \cite{3}.

\par
Another context, intimately connected with the previous one, is that of
 parasupersymmetric quantum mechanics (PSSQM) essentially developed at the
 start by Rubakov-Spiridonov \cite{4} and Beckers-Debergh \cite{5}, this field
 subtending now the idea of superposition of (once again) usual bosons but
 with parafermions, i.e. ``particles" characterized by specific trilinear
 structure relations associated with parastatistics of order $p$ \cite{6,7}.
 Let me only recall that, for $p = 1$, the parafermionic context reduces 
to the above fermionic one, such a remark justifying that, in the following 
PSSQM developments, I want to consider $p \geq 2$. PSSQM has, in particular,
 specific interests in the study of interactions for real particles (i.e. when
 $p = 2$, nonzero rest mass vector mesons in external (electro)magnetic 
fields \cite{8} have been studied).

\par
Both contexts can then be visited in connection with recent developments in
 quantum deformations \cite{9} and, more particularly, in 
$\underline{\hbox{generalized}}$ quantum deformations according to specific 
points of view \cite{10}. Let me
 say that, at the levels  of $q$-deformed SSQM and PSSQM, there are already
 published works \cite{11,12} which will not enter into the present discussion.
 I mainly want to insist here on the interest of generalized quantum
 deformations leading to recent results in SSQM as well as in PSSQM. The
 contents of this communication are then distributed as follows. In Section 2,
 I recall a few necessary ingredients and, in Sections 3 and 4, I successively
 consider the SSQM- and PSSQM-contexts and some of their generalized
 deformations. In the parasupersymmetric case, I limit myself to the $p 3D
 2$-study
 while the general one for arbitrary $p$'s is already available \cite{13}.

\section{A few necessary ingredients}

 For evident reasons, it is not possible to insert here the complete
 set of ingredients covering SSQM- and PSSQM-characteristics, but I can men
tion useful references for necessary information besides a few relations.

\par
Let me recall that $N = 2$-SSQM is, after Witten \cite{1}, characterized 
by a
 Lie superalgebra \cite{14} called sqm(2) generated by two supercharges 
$Q_1$ and $Q_2$ (or $Q = {1 \over\sqrt{2}} (Q_1 + i Q_2)$ and 
$Q^+$) such that
\begin{equation}
 \{ Q,Q^+\} = H_{SS},\qquad [Q,H_{SS}] =0,\qquad [Q^+,H_{SS}] =0 ,
 \; \; \; \; Q^2 = {Q^+}^2 = 0
\end{equation}
where $H_{SS}$ appears as the supersymmetric Hamiltonian made of a bosonic 
as well as a fermionic parts, the supercharges being expressed in terms of a
 superpotential $W (x)$ (if I consider and limit myself to $1$-dimensional
 problems). It is not possible to $q$-deform such a structure \cite{11} in
 contradistinction with Spiridonov's claim, but it is possible \cite{15} 
through generalized deformations as shown in Section 3. An important property 
entering
 in such a construction is to learn that SSQM is also realized through the
 superposition of parabosons and parafermions of all orders $p\,(\neq 1)$ if 
we combine necessarily pairs of the $\underline{\hbox{same}}$ orders \cite{16}. Such a
 property is obtained through Green-Cusson Ansatze \cite{17}, i.e.
 when the realizations of $p = 2$-parabosonic $(a)$ and parafermionic $(b)$ 
annihilation operators are
\begin{equation}
 a = \sum_{\alpha\,= 1}^{p\,= 2} A_{\alpha}\; \xi_{\alpha},\qquad
 b = \sum_{\beta\,= 1}^2 B_{\beta}\; \xi_{\beta}
\end{equation}
where
\begin{equation}
 [A_{\alpha},A_{\beta}^+] = \delta_{\alpha \, \beta},\qquad
 \{B_{\alpha},B_{\beta}^+ \} = \delta_{\alpha \, \beta},\qquad
 \{\xi_{\alpha},\xi_{\beta} \} = 2 \delta_{\alpha \, \beta},
\end{equation}
referring to usual bosonic ($A_{\alpha},\, \alpha = 1,2$) and fermionic
 ($B_{\alpha},\, \alpha = 1,2$) operators while the $\xi$'s generate
 $\cal{C} {\ell}$$_2$-Clifford algebras. Discussions on Fock bases associated
 with operators such as (2) can be presented following Mac\-farlane's 
developments \cite{18}, so that we can enter two usual bosons ($a_1,a_2$) 
and four usual fermions ($b_1,b_2,b_3,f$) with a view to study the action 
of the eight $osp (2{\mid}2)$-symmetry operators on this space (see Refs.
\cite{15} and \cite{19}).
One gets, in correspondence with eqs.(2), that
\begin{equation}
 a = \sqrt{2}\,(a_1f + a_2f^+),\qquad
 b = \sqrt{2}\,[b_1(b_3 + b_3^+)f + ib_2(b_3 - b_3^+)f^+]
\end{equation}
with  
\begin{eqnarray}
&&  a_1 = \frac{1}{\sqrt{2}}(A_1 + iA_2),\qquad
a_2 =  \frac{1}{\sqrt{2}}(A_1 - iA_2),\qquad
 [a_1,a_2] = 0,\qquad [a_\alpha,f] = 0, \nonumber \\
&& \{f,f^+ \} = 1,\qquad [b_1,b_2] = \cdots = 0,\qquad
 [b_j,f] = 0,\qquad j = 1,2,3. \nonumber
\end{eqnarray}
The corresponding $osp (2{\mid}2)$-operators \cite{20} given by
\begin{eqnarray}
&& H_{PB} = \frac{1}{2} \{a,a^+ \},\qquad H_{PF} = \frac{1}{2} [b
,b^+],
	\qquad C_+ = \frac{1}{2} \{a,a^+ \},\qquad C_- = \frac{1}{2}
 \{a,a \},
 \nonumber \\
&& 	Q = \frac{1}{2} \{a,b \},\qquad Q^+ = \frac{1}{2} \{b^+,
a^+ \},
 \qquad S = \frac{1}{2} \{b^+,a \}, \qquad
	S^+ = \frac{1}{2} \{a^+,b \} 
\end{eqnarray}
satisfy the $\underline{\hbox{trilinear}}$ structure relations of the 
``relative 
parabosonic set" \cite{7} whose specific characteristics are
\begin{eqnarray}
&& [ \{ a,b \},a^+] = - [ \{ a^+,b \},a] = 2 b, \nonumber \\
&& \{ \{ a,b^+ \},b \} = \{ \{ a,b \},b^+ \} = 2a,\nonumber \\
&& [ \{ a,b \},a] = [ \{ a,b \},b] = 0,\nonumber \\
&& [ \{ a^+,b \},a^+] = \{ \{ a,b^+ \},b^+ \} = 0,
\end{eqnarray}
besides common ones (with the "relative parafermionic set" \cite{7}). These
$\underline{\hbox{supersymmetric}}$ developments \cite{16} when the same order(s) of
 paraquantization is (are) considered lead to three $osp(2{\mid}2)$-irreducible
 unitary representations (a typical and two atypical ones) which open the way
 to succeed in deforming SSQM (cf. Section 3).

\par
As a last ingredient, let me introduce the new structure subtended by PSSQM
 when the Beckers-Debergh approach \cite{5} is taken as the starting point. In
 $N$ = 2-PSSQM and in generalization with respect to SSQM, we are concerned
 with Lie parasuperalgebras instead of Lie superalgebras. We have introduced
 \cite{5} double commutation relations, so that the fundamental $psqm$ (2) is
 characterized by the following structure relations
\begin{eqnarray}
&& [Q,[Q^+,Q]] = Q H_{PSS},\qquad [Q^+,[Q,Q^+]] = Q^+H_{PSS},\nonumber \\
&& [Q,H_{PSS}] = [Q^+,H_{PSS}] = 0,\qquad Q^3 = {Q^+}^3 = 0,
\end{eqnarray}
where $H_{PSS}$ appears as the parasupersymmetric Hamiltonian made of a 
bosonic as well as a $p = 2$-parafermionic parts, the parasupercharges 
$Q$ and $Q^+$ being now expressed in terms of two superpotentials $W_1 (x)$ 
and $W_2 (x)$. The latter are constrained by the relation
\begin{equation}
 W_2^2 (x) + {W'}_2 (x) = W_1^2 (x) - {W'}_1 (x) + c_1
\end{equation}
where primes refer to spatial derivatives as usual.

\section{SSQM and generalized deformations}

Let me discuss very briefly \underline{two} kinds of generalized
 deformations, each of them having specific properties in connection with SSQM.

\subsection{Possible deformations of SSQM}

As already proposed elsewhere \cite{15,19}, we follow the Quesne
 suggestion \cite{10} leading to the substitution of our parabosonic and
 parafermionic operators by new deformed ones as follows:
\begin{equation}
 a \longrightarrow A = \frac{1}{2 \sqrt{2}}\,(\sqrt{F_2 (N)}\,a^2 a^+
 - \sqrt{F_1 (N)} a^+ a^2)
\end{equation}
and
\begin{equation}
 b \longrightarrow B = \frac{1}{2 \sqrt{2}} (\sqrt{F_1}\,b^+ b^2
 + \sqrt{F_2}\,b^2 b^+),
\end{equation}
where, in particular, $N$ is the number operator and $F_1, F_2$ some arbitrary
 functions. Through the construction of the operators (4) and (5) as well as
 through their actions on the (unchanged) Fock basis entering two bosons and
 four fermions, it is possible \cite{15} to see that the corresponding
 \underline{typical} irreducible representation of $osp (2{\mid}2)$ shows that
 it contains a deformed $sqm \, (2)$-representation while the other two 
atypical
ones are not deformed. Some nilpotencies are now of the third order instead 
of the second one as expected in $sqm \, (2)$.

\subsection{Reducibility of SSQM}

Here I suggest the substitution
\begin{equation}
 a \longrightarrow A = \frac{1}{2 \sqrt{2}}\,a\,(1 + P)
\end{equation}
at the level of the bosonic annihilation operator and, consequently,
\begin{equation}
 a^+ \longrightarrow A^+ = \frac{1}{2 \sqrt{2}}\,a^+\,(
1 - P)
\end{equation}
at the level of the corresponding creation operator, where $P$ is the parity
 operator admitting $(- 1)^n$ (for all integers $n$) as eigenvalues. In fact,
 these are once again \underline{deformed} operators in the sense that, 
acting on a Fock basis $\{ \mid n \rangle \}$, we get
\begin{equation}
 A \mid n \rangle = \sqrt{F (n)} \mid n - 1 \rangle,\qquad A^+ \mid n \rangle =
 \sqrt{F (n + 1)} \mid n + 1 \rangle
\end{equation}
with 
\begin{equation}
 \sqrt{F (n)} = \frac{1}{\sqrt{2}} (1 + (- 1)^n) \sqrt{n},\qquad
 \sqrt{F (n + 1)} = \frac{1}{\sqrt{2}} (1 - (- 1)^n) \sqrt{n + 1}.
\end{equation}
Such an approach is a certain generalization of the usual $q$-deformation
 introduced \cite{21}, for example, in the harmonic oscillator context.

\par
The remarkable fact here is that these operators $A$ and $A^+$ can play
 the role of ``supercharges" in SSQM. Indeed they generate the structure 
relations (1) with
\begin{equation}
 H_{SS} = \frac{1}{2} \{ A,A^+ \} = \frac{1}{2} \{ a,a^+ \}=
 - \frac{1}{2} P
\end{equation}
where $(- \frac{1}{2} P)$ with the eigenvalues $(\mp \frac{1}{2})$ plays the
 role of the fermionic Hamiltonian. Such considerations lead to 
$\underline{\underline{exact}}$ supersymmetry \cite{22} and to an (unexpected) reducibility 
of SSQM when the superpotential characteristic of the interaction is odd (but 
not when it is even). Specific $\underline{\hbox{dynamical}}$ symmetries can also be 
displayed \cite{22} but cannot be discussed here : the harmonic 
(super)oscillator enters in the
 $\underline{\hbox{odd}}$ context while the hydrogen atom (and its superCoulomblike
 interaction) belongs to the $\hbox{\underline{even}}$ case.

\section{PSSQM and generalized deformations}

Let me come back on the superposition of a usual boson and a generalized 
deformed parafer- mion of order $p = 2$. The corresponding generalized
 deformed parafermionic operators (called here $b$ and $b^+$) transform
 under a 3-dimensional unitary and irreducible representation of a 
Polychronakos-Ro${\breve{\rm c}}$ek deformed $su \,(2)$-algebra \cite{10}. 
From the parastatistical
 point of view, we are asking for generalized deformed parafermionic operators
 satisfying the nilpotency and trilinear relations
$$
 b^3 = (b^+)^3 = 0,\qquad [b,[b^+,b]] = G (N) \, b,
$$
\begin{equation}
 [b^+,[b,b^+]] = b^+ G (N),\qquad G (N) = 2 F(N + 1)
 - F (N) - F (N + 2),
\end{equation}
$$
 F (N) = b^+ b, \nonumber
$$
F being any positive analytic function. We have constructed \cite{13} 
associated parasupercharges leading to a deformed parasuperalgebra containing 
explicitly the new parasuperhamiltonian in correspondence with the structure 
\cite{7}. Through the constraints (8) on the superpotentials, we can determine 
the diagonal parasupersymmetric elements of $H_{PSS}$ as given by
\begin{equation}
 H_{kk} = \frac{1}{2} p^2 + f_k (x),\qquad k = 1,2,3,
\end{equation}
leading to only \underline{three} different Hamiltonians $H^{(1)}, H^{(2)}$
 and $H^{(3)}$. It has to be noticed that $H^{(1)}$ and $H^{(2)}$ appear as
 of the $\Xi$-type in the Semenov-Chumakov scheme \cite{23} while $H^{(3)}$
 is of the $V$-type. These specific results correspond to 3-level systems 
of special physical interest in quantum optics \cite{24}.

\section{Acknowledgment}

Thanks are due to my collaborator Dr. Nathalie Debergh for, in 
particular, a careful reading of this manuscript.

\end{document}